%% file: CERN_preprint.tex
\documentclass[ALICE,manyauthors]{cernphprep}
\usepackage{graphicx}
\usepackage{color}
\usepackage{lineno}
\usepackage{hyperref}
\usepackage{subfigure}
\usepackage{amsmath}
\usepackage{tabularx}
\usepackage{adjustbox}
\usepackage{booktabs}
\usepackage{rotating}
\usepackage{pdflscape}

\RequirePackage[normalem]{ulem} 
\RequirePackage{color}\definecolor{RED}{rgb}{1,0,0}\definecolor{BLUE}{rgb}{0,0,1} 

\newcommand{\jpsi}{\mathrm{J/}\psi}

\begin{document}
\begin{titlepage}
\PHyear{2014}
\PHnumber{149}                 
\PHdate{22 June}              
\title{Exclusive $\jpsi$ photoproduction off protons\\ in ultra-peripheral p--Pb collisions at $\mathbf{\sqrt{\textit{s}_{\rm NN}}}$ =5.02 TeV}
\ShortTitle{Exclusive J/$\mathbf{\psi}$ photoproduction off protons in ultra-peripheral p--Pb collisions}   
\Collaboration{ALICE Collaboration%
         \thanks{See Appendix~\ref{app:collab} for the list of collaboration
                      members}}
\ShortAuthor{ALICE Collaboration}      
\begin{abstract}
We present the first measurement at the LHC of exclusive J/$\psi$ photoproduction off protons, in ultra-peripheral proton-lead collisions at $\sqrt{s_{\rm NN}}=5.02$ TeV. Events are selected with a dimuon pair produced either in the rapidity interval, in the laboratory frame, $2.5<y<4$ (p--Pb) or $-3.6<y<-2.6$ (Pb--p), and no other particles observed in the ALICE acceptance. The measured cross sections $\sigma (\gamma + {\rm p} \rightarrow \jpsi + {\rm p})$ are 33.2 $\pm$ 2.2 (stat) $\pm$ 3.2 (syst) $\pm$ 0.7 (theo) nb in p--Pb and 284 $\pm$ 36 (stat) $^{+27}_{-32}$ (syst) $\pm$ 26 (theo) nb in Pb--p collisions. We measure this process up to about 700 GeV in the $\gamma {\rm p}$ centre-of-mass, which is a factor of two larger than the highest energy studied at HERA. The data are consistent with a power law dependence of the $\jpsi$ photoproduction cross section in $\gamma {\rm p}$ energies from about 20 to 700 GeV, or equivalently, from Bjorken-$x$ between $\sim 2\times 10^{-2}$ to $\sim 2\times 10^{-5}$, thus indicating no significant change in the gluon density behaviour of the proton between HERA and LHC energies.
\end{abstract}
\end{titlepage}
\setcounter{page}{2}
\input{text}  
%
\newenvironment{acknowledgement}{\relax}{\relax}
\begin{acknowledgement}
\section*{Acknowledgements}
\input{acknowledgements_march2013.tex}    
\end{acknowledgement}

\input{biblio}              
\newpage
%
%
\appendix
\section{The ALICE Collaboration}
\label{app:collab}
\input{Alice_Authorlist_2014-May-12-CERNPREP}  
\end{document}

%% file: text.tex
Exclusive $\jpsi$ photoproduction off protons is defined by a reaction in which the $\jpsi$ is produced from a $\gamma \rm{p}$ interaction, where the proton emerges intact: $\gamma +  \mathrm{p}\rightarrow \jpsi + \mathrm{p}$. This process allows a detailed study of the gluon distribution in the proton, since its cross section is expected to scale as the square of the gluon probability density function (PDF), according to leading order QCD calculations~\cite{Ryskin:1992ui}. The mass of the charm quark provides an energy scale large enough to allow perturbative QCD calculations, albeit with some theoretical uncertainties~\cite{Armesto:2014sma}. This process provides a powerful tool to search for gluon saturation~\cite{Gribov:1984tu,Mueller:1989st}, which is the most straightforward mechanism to slow down the growth of the PDF for gluons carrying a small fraction of the momentum of hadrons (Bjorken-$x$). Finding evidence of gluon saturation has become a central task for present experiments and for future projects~\cite{AbelleiraFernandez:2012ni,Accardi:2012qut} that aim to study Quantum Chromo-Dynamics (QCD). 

Both ZEUS and H1 measured the exclusive $\jpsi$ photoproduction off protons at $\gamma {\rm p}$ centre-of-mass energies ranging from 20 to 305 GeV~\cite{Chekanov:2002xi,Aktas:2005xu,Alexa:2013xxa}. This process has also been studied in pp~\cite{Aaij:2014iea}, p$\bar{\rm p}~$\cite{Aaltonen:2009kg} and heavy-ion collisions~\cite{Abelev:2012ba,Abbas:2013oua,Afanasiev:2009hy}.

In this Letter we present the first measurement of exclusive $\jpsi$ photoproduction in collisions of protons with Pb nuclei at centre-of-mass energy per nucleon pair $\sqrt{s_{\rm NN}} = 5.02$ TeV. The $\jpsi$ is produced by the interaction of a photon with either a proton or a nuclear target, where the photon is emitted from one of the two colliding particles. Although both $\gamma + \mathrm{p}\rightarrow \jpsi + \mathrm{p}$ and $\gamma + \mathrm{Pb} \rightarrow \jpsi + \mathrm{Pb}$ can occur, the Pb electric charge makes the photon emission by ion to be strongly enhanced with respect to that from the proton~\cite{Frankfurt:2006tp,Guzey:2013taa}.

The main ALICE detector used in this analysis is the single-arm muon spectrometer~\cite{Aamodt:2008zz}, covering the pseudorapidity interval $-4.0<\eta<-2.5$. The beam directions of the LHC were reversed in order to measure both forward and backward rapidity. Thus, $\jpsi$s are reconstructed in the $2.5<y<4.0$ (p--Pb) and $-3.6<y<-2.6$ (Pb--p) rapidity intervals, where $y$ is measured in the laboratory frame with respect to the proton beam direction\footnotemark \footnotetext{The ALICE detector acceptance is given in the laboratory pseudorapidity $\eta$. The convention in ALICE is that the muon spectrometer is located at $\eta<0$. In contrast, the laboratory rapidity $y$ will change sign according to the proton beam direction, from which it takes its orientation. In p-Pb, for example, the proton goes in the $\eta<0$ direction, and $y>0$.}. The $\gamma \mathrm{p}$ centre-of-mass energy $W_{\gamma {\rm p}}$ is determined by the $\jpsi$ rapidity: $W_{\gamma {\rm p}}^2 = 2 E_p M_{\jpsi} \exp (-y)$, where $M_{\jpsi}$ is the $\jpsi$ mass, $y$ is the $\jpsi$ rapidity and $E_p$ is the proton energy ($E_p =  4$ TeV in the lab frame), while the Bjorken-$x$ is given by  $x = (M_{\jpsi} /W_{\gamma {\rm p}})^2$. We study $21<W_{\gamma {\rm p}}<45$ GeV for $y>0$ and $577< W_{\gamma {\rm p}}<952$ GeV for $y<0$, thereby exceeding the $\gamma \rm{p}$ range of HERA.  

The muon spectrometer consists of a ten interaction length absorber, followed by five tracking stations, each made of two planes of cathode pad chambers, with the third station placed inside a dipole magnet with a 3 T$\cdot$m integrated magnetic field. The muon trigger system, downstream of the tracking chambers, consists of  four planes of resistive plate chambers placed behind a 7.2 interaction length iron wall. The single muon trigger threshold for the data used in this analysis was set to transverse momentum $p_{\mathrm{T}}$ = 0.5 GeV/$c$.  Other detectors used in this analysis are the Silicon Pixel Detector (SPD), VZERO and Zero Degree Calorimeters (ZDC)~\cite{Aamodt:2008zz}. The central region $|\eta|<$ 1.4 is covered by the SPD consisting of two cylindrical layers of silicon pixels. The pseudorapidity interval $2.8< \eta <5.1$ is covered by VZERO-A and $-3.7<\eta <-1.7$ by VZERO-C. These detectors are scintillator tile arrays with a time resolution better than 1 ns, allowing us to distinguish between beam-beam and beam-gas interactions. The two ZDCs are located at $\pm$112.5 m from the interaction point, and are used to detect neutrons and protons emitted in the very forward region.

The trigger for the p--Pb configuration required two oppositely charged tracks  in the muon spectrometer, and a veto on VZERO-A beam-beam interactions. In the Pb--p configuration, the trigger purity was improved with respect to the p--Pb by suppressing beam-induced backgrounds. This was achieved by requiring at least one hit in VZERO-C beam-beam trigger and a veto on VZERO-A beam-gas trigger. The integrated luminosity $L$ was corrected for the probability that exclusivity requirements could be spoiled by multiple interactions in the same bunch crossing. This pile-up correction is on average 5\%, giving $L$ = 3.9 nb$^{-1}$ $\pm$ 3.7\% (syst) for p--Pb and $L$ = 4.5 nb$^{-1}$ $\pm$ 3.4\% (syst) for Pb--p data~\cite{Abelev:2014epa}. 

Events with exactly two reconstructed tracks in the muon spectrometer were selected offline. The muon tracks had to fulfill the requirements on the radial coordinate of the track at the end of the absorber and on the extrapolation to the nominal vertex, as described in~\cite{Abelev:2012ba,Abelev:2013yxa}. Both track pseudorapidities were required to be within the chosen range $-4.0<\eta_{\mathrm{track}}<-2.5$ for p--Pb and  $-3.7<\eta_{\mathrm{track}}<-2.5$ for Pb--p. Track segments in the tracking chambers must be matched with corresponding segments in the trigger chambers. The dimuon rapidity was in the range $2.5<y<4.0$ for p--Pb and  $-3.6<y<-2.6$ for Pb--p. The chosen range in Pb--p ensured that the muon tracks are in the overlap of the muon spectrometer and VZERO-C geometrical acceptance, as VZERO-C was part of the trigger in Pb-p. A cut on VZERO timing was imposed offline to be compatible with crossing beams. In order to reduce contamination from non-exclusive $\jpsi$s that come mainly from proton dissociation, only events with no mid-rapidity tracklets (track segments formed by two hits at each SPD layer) were kept. For the same reasons, events with neutron or proton activity in any of the ZDCs were rejected.

\begin{figure}[htp]
\centering
\begin{tabular}{cc}
\includegraphics[width=1.0\columnwidth]{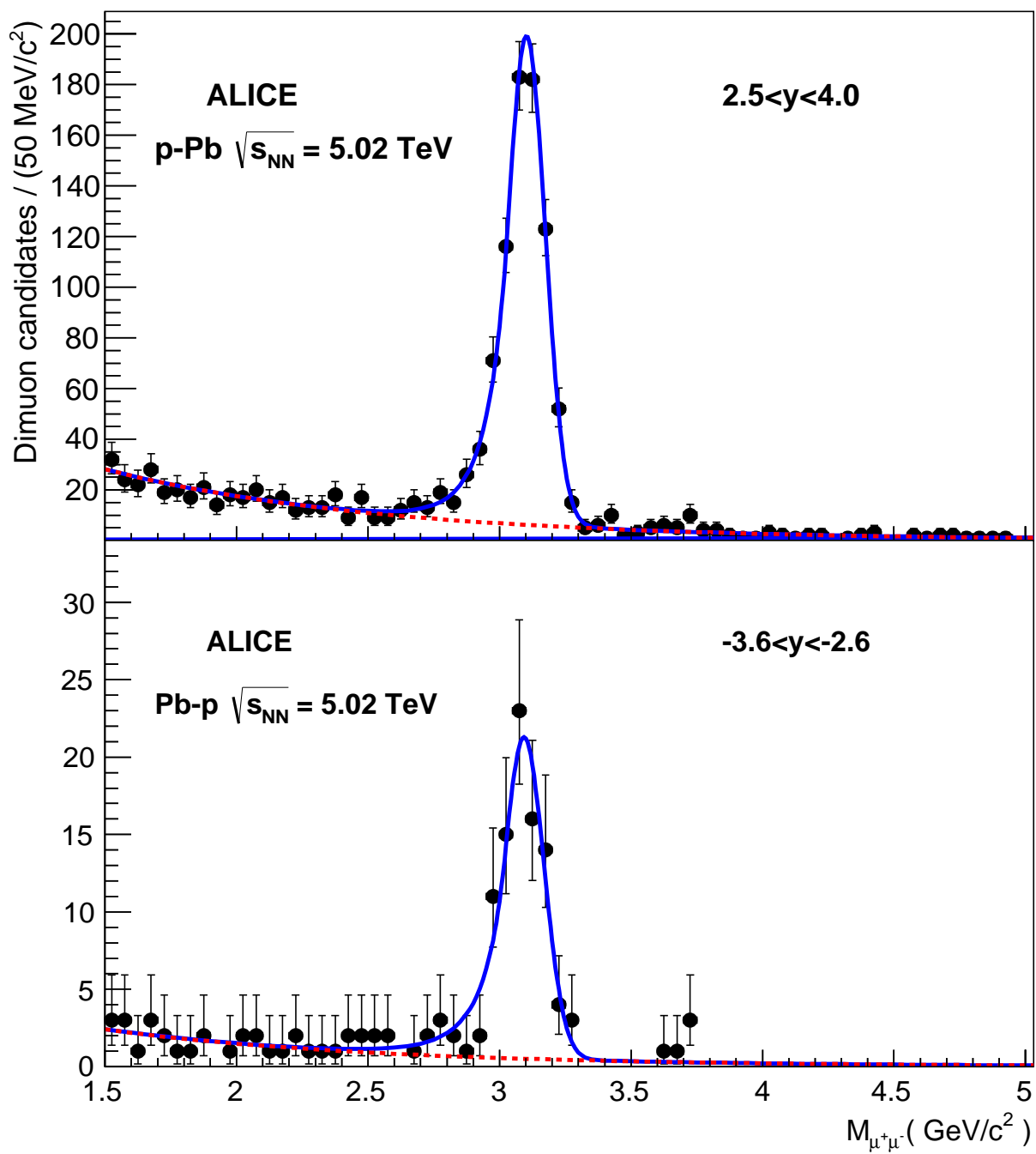}&
\end{tabular}
\caption{Invariant mass distribution for events with two oppositely charged muons, for both forward (top panel) and backward (bottom panel) dimuon rapidity samples.}
\label{figure1}
\end{figure}

The dimuon invariant mass  spectra ($M_{\mu^{+}\mu^{-}}$) after these selections are shown in Fig.~\ref{figure1}. The $\jpsi$ peak is clearly visible in both data sets, and is well described by a Crystal Ball parametrization~\cite{Gaiser:1982yw}, which yields masses and widths in agreement with the Monte Carlo simulations. The dimuon continuum is well described by an exponential as expected from two-photon production of continuum pairs ($\gamma \gamma \rightarrow \mu^{+}\mu^{-}$)~\cite{Abelev:2012ba,Abbas:2013oua}. 

\begin{figure}[htp]
\centering
\begin{tabular}{cc}
\includegraphics[width=1.0\columnwidth]{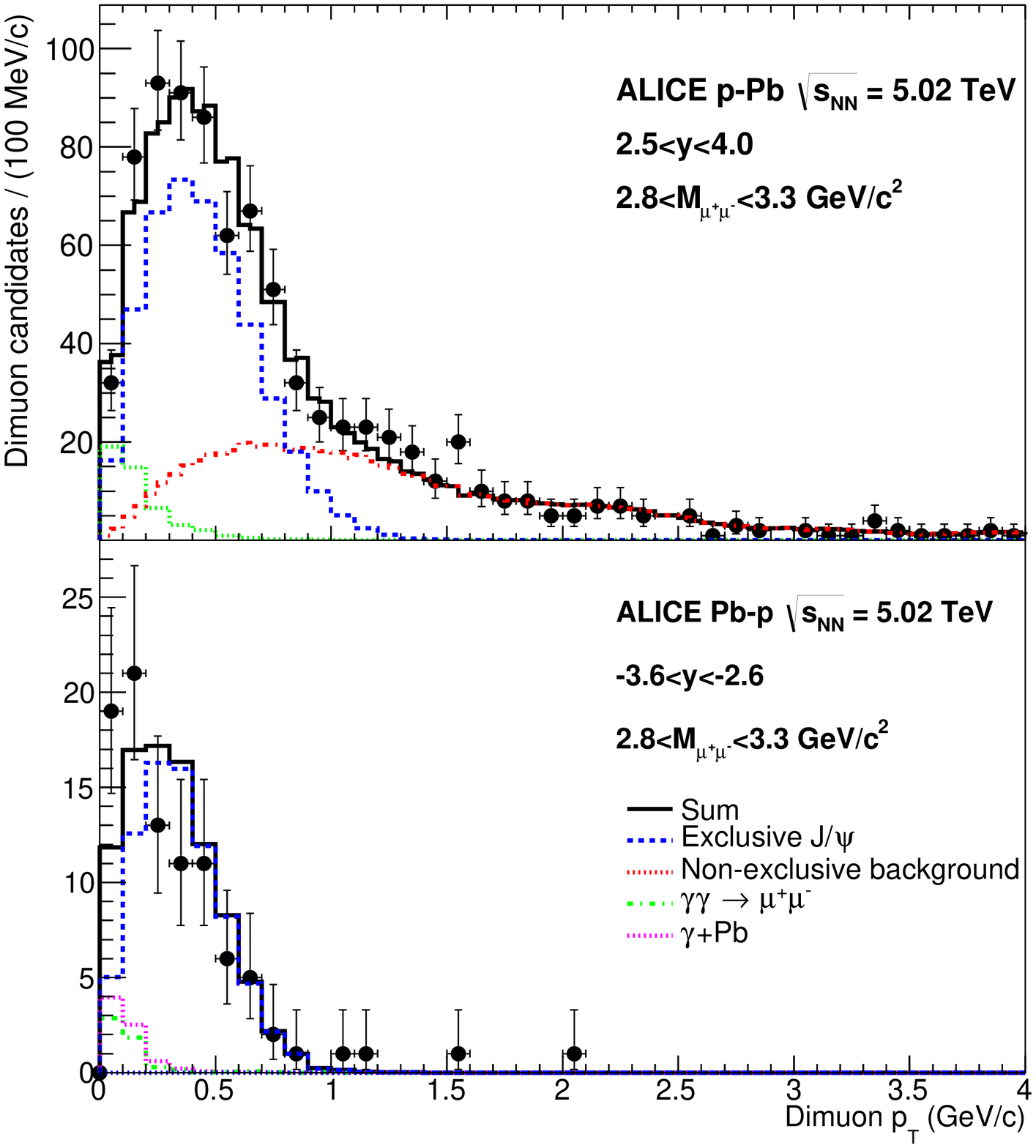}
\end{tabular}
\caption{Transverse momentum distribution for events with two oppositely charged muons, for both forward (top panel) and backward (bottom panel) dimuon rapidity samples.}
\label{figure2}
\end{figure}

The extracted number of $\jpsi$s obtained from the invariant mass fit includes a mix of exclusive and non-exclusive J/$\psi$ candidates. A different $p_{\mathrm{T}}$ distribution is expected from exclusive and non-exclusive J/$\psi$ events~\cite{Alexa:2013xxa}. For this reason, the number of exclusive J/$\psi$s can be determined from the dimuon $p_{\mathrm{T}}$ distributions shown in Fig.~\ref{figure2}. The bulk of dimuon events having $p_{\mathrm{T}}<$ 1 GeV/$c$ is mainly due to exclusive $\jpsi$ production, while the tail extending up to higher $p_{\mathrm{T}}$ on the top panel (p--Pb) comes from non-exclusive interactions. Exclusive $\jpsi$ coming from $\gamma$p interactions and $\gamma \gamma$ contribute to both $p_{\mathrm{T}}$ spectra. In addition, for p--Pb, a background, coming from non-exclusive $\jpsi$s and non-exclusive $\gamma \gamma \rightarrow \mu^{+}\mu^{-}$ events was taken into account, while for the Pb--p sample a contribution from coherent $\jpsi$ in $\gamma$Pb interactions was considered. The latter process was neglected in p--Pb as it amounts to less than 2\%~\cite{Guzey:2013taa}. If modifications to the nuclear gluon distribution, also known as nuclear shadowing, are considered this contribution would be even smaller. Here, an additional 50\% reduction is expected~\cite{Abbas:2013oua} from shadowing effects. The $p_{\mathrm{T}}$ shapes for the $\jpsi$ in $\gamma {\rm p}$, $\gamma \gamma \rightarrow \mu^{+}\mu^{-}$, and coherent $\jpsi$ in $\gamma {\rm Pb}$ components (Monte Carlo templates) were obtained using STARLIGHT~\cite{Klein:1999qj,starlight} events folded with the detector response simulation. For p--Pb, these templates were fitted to the data leaving the normalization free for $\jpsi$ in $\gamma$p and the non-exclusive background. The $\gamma \gamma \rightarrow \mu^{+}\mu^{-}$ component was constrained from the invariant mass fit shown in Fig.~\ref{figure1}~\cite{Abelev:2012ba}. The non-exclusive contributions were subtracted using this fitting procedure, giving $N_{\jpsi}$. 

The $p_{\rm T}$ distribution of non-exclusive $\jpsi$ candidates and the non-exclusive dimuon continuum were obtained from data, using the same event selection as above, but requiring events to have more than two hits in the VZERO-C counters. At HERA the ratio of the non-exclusive J/$\psi$ production cross section to the exclusive one was found to decrease with $W_{\gamma\rm{p}}$~\cite{Alexa:2013xxa}. Extrapolating, this means a factor 2 smaller non-exclusive $\jpsi$ contribution in the Pb--p sample. We note that for this sample dissociation products went towards VZERO-A, used as veto at the trigger level, providing an explanation on the negligible non-exclusive contribution observed. 

The number of exclusive $\jpsi$ coming from $\gamma {\rm p}$ interactions ($N^{\rm{exc}}_{\jpsi}$) was obtained as $N^{\mathrm{exc}}_{\jpsi} = N_{\jpsi} /({1+f_{\mathrm{D}}})$, where ${f_{\mathrm{D}}}$ is the fraction of $\jpsi$ mesons coming from the decay of $\psi$(2S). Following the procedure described in~\cite{Abelev:2012ba,Abbas:2013oua}, we obtained ${f_{\mathrm{D}}}$ = 7.9$^{+2.4}_{-1.9}$\% (syst) in p--Pb and ${f_{\mathrm{D}}}=11^{+3.6}_{-2.8}$\% (syst) in Pb--p. The contribution of exclusive $\chi_{c}$ states was neglected, as these are expected to be strongly suppressed in proton-nucleus collisions~\cite{Schramm:1996aa,HarlandLang:2010ep}. The resulting yield is $N^{\rm{exc}}_{\jpsi}$ (p--Pb) = 414 $\pm$ 28 (stat) $\pm$ 27 (syst).  

$N^{\mathrm{exc}}_{\jpsi}$ in the Pb--p sample was obtained by event counting, and then subtracting the $\gamma \gamma$ and the $\gamma {\rm Pb}$ components as well as the feed-down from $\psi$(2S) decays. Based on our recent coherent J/$\psi$ results in $\gamma {\rm Pb}$~\cite{Abelev:2012ba},  taking into account the difference in the centre-of-mass energy, we estimated that 7 $\pm$ 2 (stat) events are expected in this sample. We obtained $N^{\mathrm{exc}}_{\jpsi}$ (Pb--p) = 71 $\pm$ 9 (stat) $^{+2}_{-5}$ (syst). A compatible number for $N^{\mathrm{exc}}_{\jpsi}$ was found studying the $\jpsi$ $p_{\mathrm{T}}$ (see Fig.~\ref{figure2} bottom panel). The exclusive $\jpsi$ template was obtained by changing the exponential slope of the $p_{\mathrm{T}}^{2}$ spectrum in the Monte Carlo from its default value of 4.0 to 6.7 (GeV/$c$)$^{-2}$. This value agrees with an extrapolation of the $W_{\gamma {\rm p}}$ dependence of the $p_{\mathrm{T}}^{2}$ slope seen by H1~\cite{Alexa:2013xxa}.

The product of the detector acceptance and efficiency A$\times \varepsilon$ for $\jpsi$ was calculated using STARLIGHT and ranges from 11\% to 31\% for the rapidity intervals corresponding to the measurements given in Table~\ref{table2}. The systematic uncertainties on the measurement of the $\jpsi$ cross section are listed in Table~\ref{table1}. The cross sections  corresponding to exclusive $\jpsi$ photoproduction off protons were obtained using $\frac{\mathrm{d} \sigma}{\mathrm{d}y} = \frac{N^{\mathrm{exc}}_{\jpsi}}{(\mathrm{A} \times \varepsilon) \cdot BR \cdot L \cdot \Delta y}$, where $BR$ is the branching ratio and $\Delta y$ is the rapidity interval. We obtained $\frac{{\rm d} \sigma}{{\rm d}y}$ =  6.42 $\pm$ 0.43 (stat) $\pm$ 0.61 (syst)  $\mu$b for p--Pb and $\frac{{\rm d} \sigma}{{\rm d}y}$ = 2.46 $\pm$ 0.31 (stat) $^{+0.24}_{-0.28}$ (syst) $\mu$b for Pb--p collisions (see Table~\ref{table2}).
%
\begin{table}
\begin{center}
\begin{tabular}{lcr}
Source&p--Pb & Pb--p\\
\hline
Signal extraction & 6\% & $^{+0.0}_{-6.0}$\%\\
Luminosity~\cite{Abelev:2014epa} & 3.3\% & 3.0\% \\
Tracking efficiency~\cite{Abelev:2013yxa}& 4\% & 6\% \\
Muon Trigger efficiency~\cite{Abelev:2013yxa} & 2.8\% & 3.2\% \\
Matching & 1\% & 1\% \\
VZERO-C efficiency & - & 3.5\% \\
\hline
Total uncorrelated & 8.5\% & $^{+8.3}_{-10.2}$\% \\
\hline
Luminosity~\cite{Abelev:2014epa} & 1.6\% & 1.6\% \\
Branching ratio~\cite{Beringer:1900zz} & 1\% & 1\% \\
VZERO-A veto efficiency & $^{+2.0}_{-0.0}$\% & $^{+2.0}_{-0.0}$\% \\
Feed-down & $^{+1.8}_{-2.2}$\% & $^{+2.6}_{-3.1}$\% \\
$\jpsi$ acceptance & 3\%  & 3\% \\
\hline
Total & $\pm$9.6\% & $^{+9.6}_{-11.3}$\% \\  
\end{tabular}
\end{center}
\caption{Summary of the contributions to the systematic uncertainty for the integrated $\jpsi$ cross section measurement for the full rapidity intervals.}
\label{table1}
\end{table}

We measured the cross section for the exclusive $\gamma\gamma \rightarrow \mu^{+}\mu^{-}$ process at invariant mass $1.5<M_{\mu^{+}\mu^{-}}<2.5$ GeV/$c^{2}$ and in the rapidity range $2.5<y<4.0$, using the same technique as for the $\jpsi$ to remove the non-exclusive background, obtaining $\sigma(\gamma \gamma \rightarrow \mu^{+}\mu^{-})$ = 1.76 $\pm$ 0.12 (stat) $\pm$ 0.16 (syst) $\mu$b for this kinematic range. The STARLIGHT prediction for this standard QED process is 1.8 $\mu$b, which is in good agreement with this measurement. This provides an additional indication that the non-exclusive background subtraction is under control.

The cross section $\frac{{\rm d} \sigma}{{\rm d}y}({\rm p+Pb} \rightarrow {\rm p+Pb}+\jpsi)$ is related to the photon-proton cross section, $\sigma(\gamma+{\rm p} \rightarrow \jpsi+{\rm p}) \equiv \sigma(W_{\gamma\rm{p}})$, through the photon flux, $\frac{\mathrm{d} n}{\mathrm{d} k}$:
\begin{equation*} 
\frac{{\rm d} \sigma}{{\rm d}y}({\rm p+Pb} \rightarrow {\rm p+Pb}+\jpsi) = k \frac{\mathrm{d}n}{\mathrm{d}k}
\sigma(\gamma+{\rm p} \rightarrow \jpsi+{\rm p}).
\end{equation*}
\noindent

Here, $k$ is the photon energy, which is determined by the $\jpsi$ mass and rapidity, $k = (1/2) M_{\jpsi} \exp{(-y)}$. The average photon flux values for the different rapidity intervals were calculated using STARLIGHT and are listed in Table~\ref{table2}. The $\langle W_{\gamma {\rm p}} \rangle$ is calculated by weighting with the product of the photon spectrum and the cross section $\sigma (\gamma p)$ from STARLIGHT. The photon spectrum is calculated in impact parameter space requiring that there should be no hadronic interaction. The uncertainty in this approach is estimated by increasing/decreasing the Pb-radius with $\pm$0.5 fm, corresponding to the nuclear skin thickness and of the same order as the upper limit for the difference between the proton and neutron radius of Pb when calculating the hadronic interaction probability. This gives an uncertainty of 9\% in the photon flux for the high energy data point and 2\% at low energy (see Table~\ref{table2}). The uncertainty is larger for the high photon energies since here one is dominated by small impact parameters and thus more sensitive to the rejection of hadronic interactions with impact parameters near the Pb radius.  
%
\begin{table}
\footnotesize
\scalebox{0.9}{
\begin{tabular}{cccccc}
 Rapidity&  $\frac{{\rm d}\sigma}{{\rm d}y} (\mu b)$ & $k \frac{\mathrm{d}n}{\mathrm{d}k}$  &   $W_{\gamma {\rm p}}$ (GeV) & $\langle W_{\gamma {\rm p}} \rangle$ (GeV)
&$\sigma (\gamma + {\rm p} \rightarrow \jpsi + {\rm p}) (\mathrm{nb}) $\\ \hline
2.5$<y<$4.0  &  6.42 $\pm$ 0.43 (stat) $\pm$ 0.61 (syst) 	& 193.3 & (21,45)  & 29.8				& 33.2 $\pm$ 2.2 (stat) $\pm$ 3.2 (syst) $\pm$ 0.7 (theo)	 \\
3.5$<y<$4.0  & 5.77 $\pm$ 0.76 (stat) $\pm$ 0.58 (syst)   		& 208.9 & (21,27)  & 24.1	         & 27.6 $\pm$ 3.6 (stat) $\pm$ 2.8 (syst) $\pm$ 0.6 (theo)	\\
3.0$<y<$3.5  & 6.71 $\pm$ 0.60 (stat)  $\pm$ 0.55 (syst) 	& 193.3 & (27,35)  & 30.9          			& 34.7 $\pm$ 3.1 (stat) $\pm$ 2.9 (syst) $\pm$ 0.7 (theo)	\\
2.5$<y<3.0$  & 6.83 $\pm$ 1.0 (stat) $\pm$ 0.75 (syst)   & 177.6                   & (35,45)  & 39.6              & 38.5 $\pm$ 5.6 (stat) $\pm$ 4.2 (syst) $\pm$ 0.8 (theo)	\\
\hline
-3.6 $<y<$ -2.6  &  2.46 $\pm$ 0.31 (stat) $^{+0.24}_{-0.28}$ (syst) &  8.66 & (577,952)  & 706 	& 284 $\pm$ 36 (stat) $^{+27}_{-32}$ (syst) $\pm$ 26 (theo) 	 \\
\end{tabular}
}
\caption{Differential cross sections for  exclusive $\jpsi$ photoproduction off protons in ultra-peripheral p--Pb and Pb--p collisions at $\sqrt{s_{\rm NN}}=5.02$ TeV. The corresponding $\jpsi$ photoproduction cross sections in bins of $W_{\gamma {\rm p}}$ are also presented.}
\label{table2}
\end{table}

\begin{figure}[h]
\includegraphics[width=1\columnwidth]{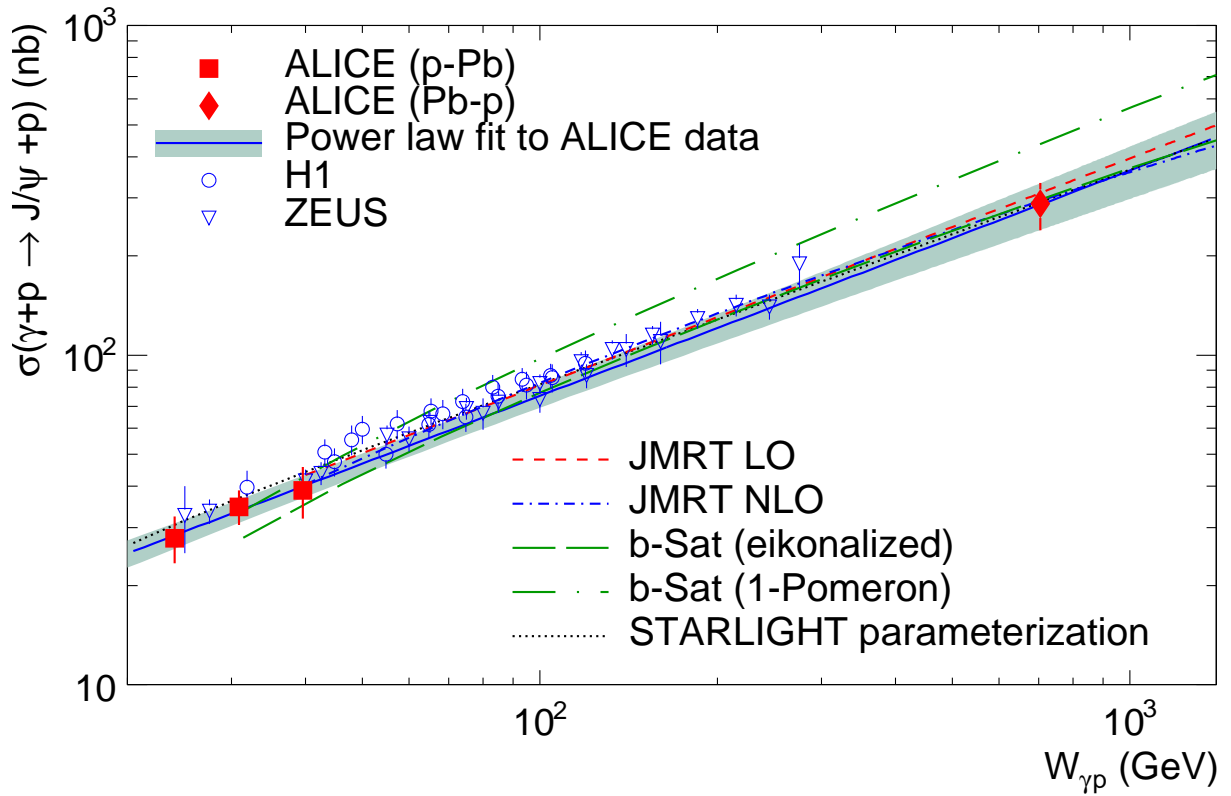}%
\caption{Exclusive $\jpsi$ photoproduction cross section off protons measured by ALICE and compared to HERA data. Comparisons to STARLIGHT, JMRT and the b-Sat models are shown. The power law fit to ALICE data is also shown.}
\label{figure3}
 \end{figure}

Figure~\ref{figure3} shows the ALICE measurements for $\sigma(W_{\gamma\rm{p}})$. Comparisons to previous measurements and to different theoretical models are also shown. As mentioned earlier, $\sigma(W_{\gamma\rm{p}})$ is proportional to the square of the gluon PDF of the proton~\cite{Ryskin:1992ui}. For HERA energies, the gluon distribution at low Bjorken-$x$ is well described by a power law in $x$~\cite{Aaron:2009aa}, which implies the cross section $\sigma(W_{\gamma\rm{p}})$ will also follow a power law. A deviation from such a trend in the measured cross section as $x$ decreases, or equivalently, as $ W_{\gamma\rm{p}}$ increases, could indicate a change in the evolution of the gluon density function, as expected at the onset of saturation. 

Both ZEUS and H1~\cite{Chekanov:2002xi,Aktas:2005xu,Alexa:2013xxa} fitted their data using a power law $\sigma \sim W_{\gamma{\rm p}}^{\delta}$, obtaining $\delta$ = 0.69 $\pm$ 0.02 (stat) $\pm$ 0.03 (syst), and $\delta$ = 0.67 $\pm$ 0.03 (stat + syst), respectively. Due to the large HERA statistics, a simultaneous fit of H1, ZEUS, ALICE low energy points data gives power-law fit parameters almost identical to those obtained from HERA alone. A fit to ALICE data alone gives $\delta$ = 0.68 $\pm$ 0.06 (stat + syst), only uncorrelated systematic errors were considered here. Thus, no deviation from a power law is observed up to about 700 GeV. 

Two calculations are available from the JMRT group~\cite{Jones:2013pga}: the first one referred to as LO is based on a power law description of the process, while the second model is labeled as NLO, and includes contributions which mimic effects expected from the dominant NLO corrections. Because both JMRT models have been fitted to the same data, the resulting energy dependences are very similar. Our data support their extracted gluon distribution up to $x\sim 2\times 10^{-5}$. The STARLIGHT parameterization is based on a power law fit using only fixed-target and HERA data, giving $\delta$ = 0.65 $\pm$ 0.02.  Figure~\ref{figure3} also shows predictions from the b-Sat eikonalized model~\cite{Kowalski:2006hc} which uses the Color Glass Condensate approach~\cite{Gelis:2010nm} to incorporate saturation, constraining it to HERA data alone. The results from the models mentioned above are within one sigma of our measurement. The b-Sat 1-Pomeron prediction taken from~\cite{AbelleiraFernandez:2012ni} also agrees with the ALICE low energy data points, but it is about 4 sigmas above our measurement at the highest energy. 

\begin{figure}[t]
\includegraphics[width=1\columnwidth]{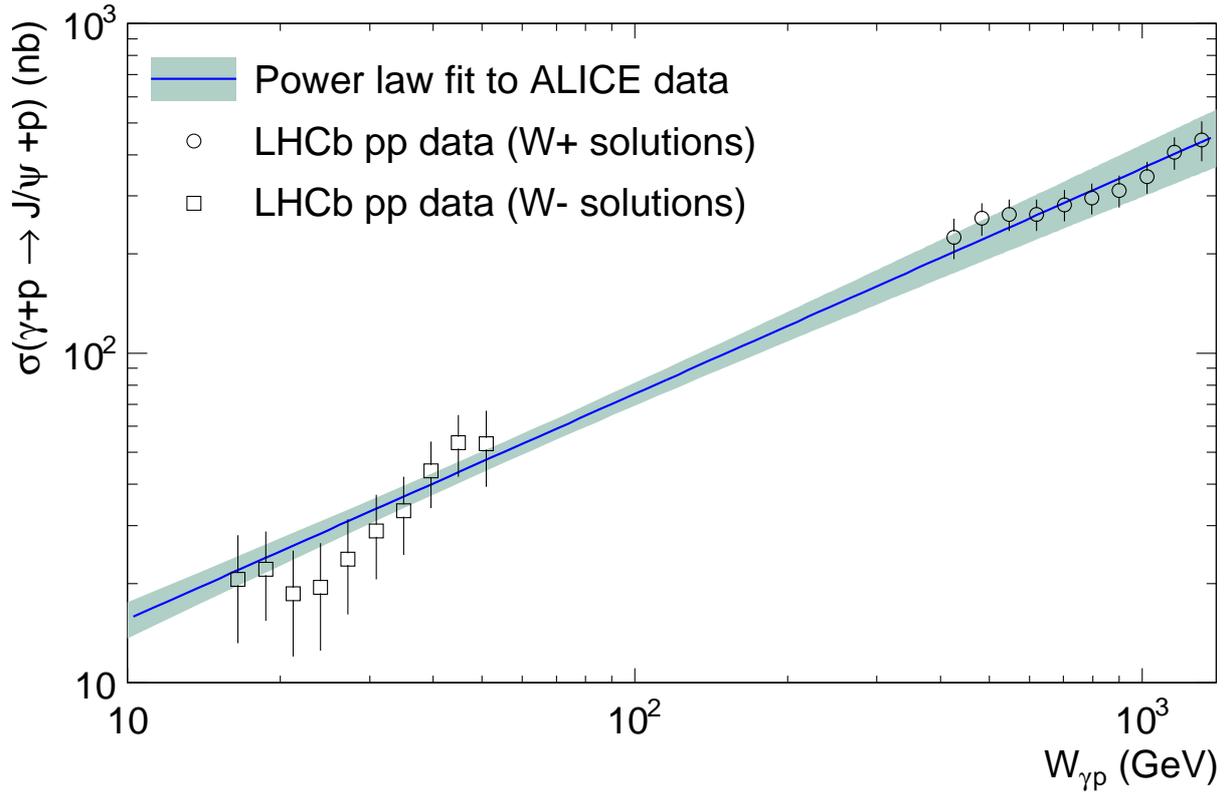}
\caption{The power law fit to ALICE data is compared to LHCb solutions.}
\label{figure4}
 \end{figure}

LHCb recently published results for $\sigma(W_{\gamma\rm{p}})$ based on exclusive $\jpsi$ production in pp collisions~\cite{Aaij:2014iea}. Their analysis, using data from a symmetric system, suffers from the intrinsic impossibility of identifying the photon emitter and the photon target. Since the non-exclusive background, as mentioned above, depends on $W_{\gamma\rm{p}}$, this feeds into the uncertainty in the subtraction of these processes, making the extraction of the underlying $\sigma(W_{\gamma\rm{p}})$ to be strongly model dependent. Moreover, in contrast with p--Pb collisions, there is a large uncertainty in the hadronic survival probability in pp collisions, as well as an unknown contribution from production through Odderon-Pomeron fusion~\cite{Aaltonen:2009kg,Schramm:1996aa}. For each $\frac{{\rm d}\sigma}{{\rm d}y}$ measurement, they reported a $\rm{W+}$ and a $\rm{W-}$ solution. These coupled solutions are shown in Figure~\ref{figure4}, together with the power law fit to ALICE measurements. Despite these ambiguities and assumptions the LHCb solutions turned out to be compatible with the power law dependence extracted from our data. 

In summary, we have made the first measurement of exclusive $\jpsi$ photoproduction off protons in p--Pb collisions at the LHC. Our data are compatible with a power law dependence of $\sigma(W_{\gamma\rm{p}})$ up to about 700 GeV in $W_{\gamma\rm{p}}$, corresponding to $x\sim 2\times 10^{-5}$. A natural explanation is that no change in the behaviour of the gluon PDF in the proton is observed between HERA and LHC energies. 

%% file: acknowledgements_march2013.tex
The ALICE Collaboration would like to thank all its engineers and technicians for their invaluable contributions to the construction of the experiment and the CERN accelerator teams for the outstanding performance of the LHC complex.
\\
The ALICE Collaboration gratefully acknowledges the resources and support provided by all Grid centres and the Worldwide LHC Computing Grid (WLCG) collaboration.
\\
The ALICE Collaboration acknowledges the following funding agencies for their support in building and
running the ALICE detector:
 \\
State Committee of Science,  World Federation of Scientists (WFS)
and Swiss Fonds Kidagan, Armenia,
 \\
Conselho Nacional de Desenvolvimento Cient\'{\i}fico e Tecnol\'{o}gico (CNPq), Financiadora de Estudos e Projetos (FINEP),
Funda\c{c}\~{a}o de Amparo \`{a} Pesquisa do Estado de S\~{a}o Paulo (FAPESP);
 \\
National Natural Science Foundation of China (NSFC), the Chinese Ministry of Education (CMOE)
and the Ministry of Science and Technology of China (MSTC);
 \\
Ministry of Education and Youth of the Czech Republic;
 \\
Danish Natural Science Research Council, the Carlsberg Foundation and the Danish National Research Foundation;
 \\
The European Research Council under the European Community's Seventh Framework Programme;
 \\
Helsinki Institute of Physics and the Academy of Finland;
 \\
French CNRS-IN2P3, the `Region Pays de Loire', `Region Alsace', `Region Auvergne' and CEA, France;
 \\
German BMBF and the Helmholtz Association;
\\
General Secretariat for Research and Technology, Ministry of
Development, Greece;
\\
Hungarian OTKA and National Office for Research and Technology (NKTH);
 \\
Department of Atomic Energy and Department of Science and Technology of the Government of India;
 \\
Istituto Nazionale di Fisica Nucleare (INFN) and Centro Fermi -
Museo Storico della Fisica e Centro Studi e Ricerche ``Enrico
Fermi", Italy;
 \\
MEXT Grant-in-Aid for Specially Promoted Research, Ja\-pan;
 \\
Joint Institute for Nuclear Research, Dubna;
 \\
National Research Foundation of Korea (NRF);
 \\
CONACYT, DGAPA, M\'{e}xico, ALFA-EC and the EPLANET Program
(European Particle Physics Latin American Network)
 \\
Stichting voor Fundamenteel Onderzoek der Materie (FOM) and the Nederlandse Organisatie voor Wetenschappelijk Onderzoek (NWO), Netherlands;
 \\
Research Council of Norway (NFR);
 \\
Polish Ministry of Science and Higher Education;
 \\
National Science Centre, Poland;
 \\
 Ministry of National Education/Institute for Atomic Physics and CNCS-UEFISCDI - Romania;
 \\
Ministry of Education and Science of Russian Federation, Russian
Academy of Sciences, Russian Federal Agency of Atomic Energy,
Russian Federal Agency for Science and Innovations and The Russian
Foundation for Basic Research;
 \\
Ministry of Education of Slovakia;
 \\
Department of Science and Technology, South Africa;
 \\
CIEMAT, EELA, Ministerio de Econom\'{i}a y Competitividad (MINECO) of Spain, Xunta de Galicia (Conseller\'{\i}a de Educaci\'{o}n),
CEA\-DEN, Cubaenerg\'{\i}a, Cuba, and IAEA (International Atomic Energy Agency);
 \\
Swedish Research Council (VR) and Knut $\&$ Alice Wallenberg
Foundation (KAW);
 \\
Ukraine Ministry of Education and Science;
 \\
United Kingdom Science and Technology Facilities Council (STFC);
 \\
The United States Department of Energy, the United States National
Science Foundation, the State of Texas, and the State of Ohio;
\\
Ministry of Science, Education and Sports of Croatia and  Unity through Knowledge Fund, Croatia.

%% file: Alice_Authorlist_2014-May-12-CERNPREP.tex


\begingroup
\small
\begin{flushleft}
B.~Abelev\Irefn{org71}\And
J.~Adam\Irefn{org37}\And
D.~Adamov\'{a}\Irefn{org79}\And
M.M.~Aggarwal\Irefn{org83}\And
G.~Aglieri~Rinella\Irefn{org34}\And
M.~Agnello\Irefn{org107}\textsuperscript{,}\Irefn{org90}\And
A.~Agostinelli\Irefn{org26}\And
N.~Agrawal\Irefn{org44}\And
Z.~Ahammed\Irefn{org126}\And
N.~Ahmad\Irefn{org18}
% ...
\renewcommand\labelenumi{\textsuperscript{\theenumi}~}

\section*{Affiliation notes}
\renewcommand\theenumi{\roman{enumi}}
\begin{Authlist}
\item \Adef{0}Deceased
\item \Adef{idp1126752}{Also at: St. Petersburg State Polytechnical University}
\item \Adef{idp3049600}{Also at: Department of Applied Physics, Aligarh Muslim University, Aligarh, India}
\item \Adef{idp3730224}{Also at: M.V. Lomonosov Moscow State University, D.V. Skobeltsyn Institute of Nuclear Physics, Moscow, Russia}
\item \Adef{idp3979312}{Also at: University of Belgrade, Faculty of Physics and "Vin\v{c}a" Institute of Nuclear Sciences, Belgrade, Serbia}
\item \Adef{idp4291840}{Permanent Address: Permanent Address: Konkuk University, Seoul, Korea}
\item \Adef{idp4843104}{Also at: Institute of Theoretical Physics, University of Wroclaw, Wroclaw, Poland}
\item \Adef{idp5763712}{Also at: University of Kansas, Lawrence, KS, United States}
\end{Authlist}

\section*{Collaboration Institutes}
\renewcommand\theenumi{\arabic{enumi}~}
\begin{Authlist}

\item \Idef{org1}A.I. Alikhanyan National Science Laboratory (Yerevan Physics Institute) Foundation, Yerevan, Armenia
\item \Idef{org2}Benem\'{e}rita Universidad Aut\'{o}noma de Puebla, Puebla, Mexico
\item \Idef{org3}Bogolyubov Institute for Theoretical Physics, Kiev, Ukraine
\item \Idef{org4}Bose Institute, Department of Physics and Centre for Astroparticle Physics and Space Science (CAPSS), Kolkata, India
\item \Idef{org5}Budker Institute for Nuclear Physics, Novosibirsk, Russia
\item \Idef{org6}California Polytechnic State University, San Luis Obispo, CA, United States
\item \Idef{org7}Central China Normal University, Wuhan, China
\item \Idef{org8}Centre de Calcul de l'IN2P3, Villeurbanne, France
\item \Idef{org9}Centro de Aplicaciones Tecnol\'{o}gicas y Desarrollo Nuclear (CEADEN), Havana, Cuba
\item \Idef{org10}Centro de Investigaciones Energ\'{e}ticas Medioambientales y Tecnol\'{o}gicas (CIEMAT), Madrid, Spain
\item \Idef{org11}Centro de Investigaci\'{o}n y de Estudios Avanzados (CINVESTAV), Mexico City and M\'{e}rida, Mexico
\item \Idef{org12}Centro Fermi - Museo Storico della Fisica e Centro Studi e Ricerche ``Enrico Fermi'', Rome, Italy
\item \Idef{org13}Chicago State University, Chicago, USA
\item \Idef{org14}Commissariat \`{a} l'Energie Atomique, IRFU, Saclay, France
\item \Idef{org15}COMSATS Institute of Information Technology (CIIT), Islamabad, Pakistan
\item \Idef{org16}Departamento de F\'{\i}sica de Part\'{\i}culas and IGFAE, Universidad de Santiago de Compostela, Santiago de Compostela, Spain
\item \Idef{org17}Department of Physics and Technology, University of Bergen, Bergen, Norway
\item \Idef{org18}Department of Physics, Aligarh Muslim University, Aligarh, India
\item \Idef{org19}Department of Physics, Ohio State University, Columbus, OH, United States
\item \Idef{org20}Department of Physics, Sejong University, Seoul, South Korea
\item \Idef{org21}Department of Physics, University of Oslo, Oslo, Norway
\item \Idef{org22}Dipartimento di Fisica dell'Universit\`{a} 'La Sapienza' and Sezione INFN Rome, Italy
\item \Idef{org23}Dipartimento di Fisica dell'Universit\`{a} and Sezione INFN, Cagliari, Italy
\item \Idef{org24}Dipartimento di Fisica dell'Universit\`{a} and Sezione INFN, Trieste, Italy
\item \Idef{org25}Dipartimento di Fisica dell'Universit\`{a} and Sezione INFN, Turin, Italy
\item \Idef{org26}Dipartimento di Fisica e Astronomia dell'Universit\`{a} and Sezione INFN, Bologna, Italy
\item \Idef{org27}Dipartimento di Fisica e Astronomia dell'Universit\`{a} and Sezione INFN, Catania, Italy
\item \Idef{org28}Dipartimento di Fisica e Astronomia dell'Universit\`{a} and Sezione INFN, Padova, Italy
\item \Idef{org29}Dipartimento di Fisica `E.R.~Caianiello' dell'Universit\`{a} and Gruppo Collegato INFN, Salerno, Italy
\item \Idef{org30}Dipartimento di Scienze e Innovazione Tecnologica dell'Universit\`{a} del  Piemonte Orientale and Gruppo Collegato INFN, Alessandria, Italy
\item \Idef{org31}Dipartimento Interateneo di Fisica `M.~Merlin' and Sezione INFN, Bari, Italy
\item \Idef{org32}Division of Experimental High Energy Physics, University of Lund, Lund, Sweden
\item \Idef{org33}Eberhard Karls Universit\"{a}t T\"{u}bingen, T\"{u}bingen, Germany
\item \Idef{org34}European Organization for Nuclear Research (CERN), Geneva, Switzerland
\item \Idef{org35}Faculty of Engineering, Bergen University College, Bergen, Norway
\item \Idef{org36}Faculty of Mathematics, Physics and Informatics, Comenius University, Bratislava, Slovakia
\item \Idef{org37}Faculty of Nuclear Sciences and Physical Engineering, Czech Technical University in Prague, Prague, Czech Republic
\item \Idef{org38}Faculty of Science, P.J.~\v{S}af\'{a}rik University, Ko\v{s}ice, Slovakia
\item \Idef{org39}Frankfurt Institute for Advanced Studies, Johann Wolfgang Goethe-Universit\"{a}t Frankfurt, Frankfurt, Germany
\item \Idef{org40}Gangneung-Wonju National University, Gangneung, South Korea
\item \Idef{org41}Gauhati University, Department of Physics, Guwahati, India
\item \Idef{org42}Helsinki Institute of Physics (HIP), Helsinki, Finland
\item \Idef{org43}Hiroshima University, Hiroshima, Japan
\item \Idef{org44}Indian Institute of Technology Bombay (IIT), Mumbai, India
\item \Idef{org45}Indian Institute of Technology Indore, Indore (IITI), India
\item \Idef{org46}Inha University, Incheon, South Korea
\item \Idef{org47}Institut de Physique Nucl\'eaire d'Orsay (IPNO), Universit\'e Paris-Sud, CNRS-IN2P3, Orsay, France
\item \Idef{org48}Institut f\"{u}r Informatik, Johann Wolfgang Goethe-Universit\"{a}t Frankfurt, Frankfurt, Germany
\item \Idef{org49}Institut f\"{u}r Kernphysik, Johann Wolfgang Goethe-Universit\"{a}t Frankfurt, Frankfurt, Germany
\item \Idef{org50}Institut f\"{u}r Kernphysik, Westf\"{a}lische Wilhelms-Universit\"{a}t M\"{u}nster, M\"{u}nster, Germany
\item \Idef{org51}Institut Pluridisciplinaire Hubert Curien (IPHC), Universit\'{e} de Strasbourg, CNRS-IN2P3, Strasbourg, France
\item \Idef{org52}Institute for Nuclear Research, Academy of Sciences, Moscow, Russia
\item \Idef{org53}Institute for Subatomic Physics of Utrecht University, Utrecht, Netherlands
\item \Idef{org54}Institute for Theoretical and Experimental Physics, Moscow, Russia
\item \Idef{org55}Institute of Experimental Physics, Slovak Academy of Sciences, Ko\v{s}ice, Slovakia
\item \Idef{org56}Institute of Physics, Academy of Sciences of the Czech Republic, Prague, Czech Republic
\item \Idef{org57}Institute of Physics, Bhubaneswar, India
\item \Idef{org58}Institute of Space Science (ISS), Bucharest, Romania
\item \Idef{org59}Instituto de Ciencias Nucleares, Universidad Nacional Aut\'{o}noma de M\'{e}xico, Mexico City, Mexico
\item \Idef{org60}Instituto de F\'{\i}sica, Universidad Nacional Aut\'{o}noma de M\'{e}xico, Mexico City, Mexico
\item \Idef{org61}iThemba LABS, National Research Foundation, Somerset West, South Africa
\item \Idef{org62}Joint Institute for Nuclear Research (JINR), Dubna, Russia
\item \Idef{org63}Konkuk University, Seoul, South Korea
\item \Idef{org64}Korea Institute of Science and Technology Information, Daejeon, South Korea
\item \Idef{org65}KTO Karatay University, Konya, Turkey
\item \Idef{org66}Laboratoire de Physique Corpusculaire (LPC), Clermont Universit\'{e}, Universit\'{e} Blaise Pascal, CNRS--IN2P3, Clermont-Ferrand, France
\item \Idef{org67}Laboratoire de Physique Subatomique et de Cosmologie, Universit\'{e} Grenoble-Alpes, CNRS-IN2P3, Grenoble, France
\item \Idef{org68}Laboratori Nazionali di Frascati, INFN, Frascati, Italy
\item \Idef{org69}Laboratori Nazionali di Legnaro, INFN, Legnaro, Italy
\item \Idef{org70}Lawrence Berkeley National Laboratory, Berkeley, CA, United States
\item \Idef{org71}Lawrence Livermore National Laboratory, Livermore, CA, United States
\item \Idef{org72}Moscow Engineering Physics Institute, Moscow, Russia
\item \Idef{org73}National Centre for Nuclear Studies, Warsaw, Poland
\item \Idef{org74}National Institute for Physics and Nuclear Engineering, Bucharest, Romania
\item \Idef{org75}National Institute of Science Education and Research, Bhubaneswar, India
\item \Idef{org76}Niels Bohr Institute, University of Copenhagen, Copenhagen, Denmark
\item \Idef{org77}Nikhef, National Institute for Subatomic Physics, Amsterdam, Netherlands
\item \Idef{org78}Nuclear Physics Group, STFC Daresbury Laboratory, Daresbury, United Kingdom
\item \Idef{org79}Nuclear Physics Institute, Academy of Sciences of the Czech Republic, \v{R}e\v{z} u Prahy, Czech Republic
\item \Idef{org80}Oak Ridge National Laboratory, Oak Ridge, TN, United States
\item \Idef{org81}Petersburg Nuclear Physics Institute, Gatchina, Russia
\item \Idef{org82}Physics Department, Creighton University, Omaha, NE, United States
\item \Idef{org83}Physics Department, Panjab University, Chandigarh, India
\item \Idef{org84}Physics Department, University of Athens, Athens, Greece
\item \Idef{org85}Physics Department, University of Cape Town, Cape Town, South Africa
\item \Idef{org86}Physics Department, University of Jammu, Jammu, India
\item \Idef{org87}Physics Department, University of Rajasthan, Jaipur, India
\item \Idef{org88}Physik Department, Technische Universit\"{a}t M\"{u}nchen, Munich, Germany
\item \Idef{org89}Physikalisches Institut, Ruprecht-Karls-Universit\"{a}t Heidelberg, Heidelberg, Germany
\item \Idef{org90}Politecnico di Torino, Turin, Italy
\item \Idef{org91}Purdue University, West Lafayette, IN, United States
\item \Idef{org92}Pusan National University, Pusan, South Korea
\item \Idef{org93}Research Division and ExtreMe Matter Institute EMMI, GSI Helmholtzzentrum f\"ur Schwerionenforschung, Darmstadt, Germany
\item \Idef{org94}Rudjer Bo\v{s}kovi\'{c} Institute, Zagreb, Croatia
\item \Idef{org95}Russian Federal Nuclear Center (VNIIEF), Sarov, Russia
\item \Idef{org96}Russian Research Centre Kurchatov Institute, Moscow, Russia
\item \Idef{org97}Saha Institute of Nuclear Physics, Kolkata, India
\item \Idef{org98}School of Physics and Astronomy, University of Birmingham, Birmingham, United Kingdom
\item \Idef{org99}Secci\'{o}n F\'{\i}sica, Departamento de Ciencias, Pontificia Universidad Cat\'{o}lica del Per\'{u}, Lima, Peru
\item \Idef{org100}Sezione INFN, Bari, Italy
\item \Idef{org101}Sezione INFN, Bologna, Italy
\item \Idef{org102}Sezione INFN, Cagliari, Italy
\item \Idef{org103}Sezione INFN, Catania, Italy
\item \Idef{org104}Sezione INFN, Padova, Italy
\item \Idef{org105}Sezione INFN, Rome, Italy
\item \Idef{org106}Sezione INFN, Trieste, Italy
\item \Idef{org107}Sezione INFN, Turin, Italy
\item \Idef{org108}SSC IHEP of NRC Kurchatov institute, Protvino, Russia
\item \Idef{org109}SUBATECH, Ecole des Mines de Nantes, Universit\'{e} de Nantes, CNRS-IN2P3, Nantes, France
\item \Idef{org110}Suranaree University of Technology, Nakhon Ratchasima, Thailand
\item \Idef{org111}Technical University of Split FESB, Split, Croatia
\item \Idef{org112}The Henryk Niewodniczanski Institute of Nuclear Physics, Polish Academy of Sciences, Cracow, Poland
\item \Idef{org113}The University of Texas at Austin, Physics Department, Austin, TX, USA
\item \Idef{org114}Universidad Aut\'{o}noma de Sinaloa, Culiac\'{a}n, Mexico
\item \Idef{org115}Universidade de S\~{a}o Paulo (USP), S\~{a}o Paulo, Brazil
\item \Idef{org116}Universidade Estadual de Campinas (UNICAMP), Campinas, Brazil
\item \Idef{org117}University of Houston, Houston, TX, United States
\item \Idef{org118}University of Jyv\"{a}skyl\"{a}, Jyv\"{a}skyl\"{a}, Finland
\item \Idef{org119}University of Liverpool, Liverpool, United Kingdom
\item \Idef{org120}University of Tennessee, Knoxville, TN, United States
\item \Idef{org121}University of Tokyo, Tokyo, Japan
\item \Idef{org122}University of Tsukuba, Tsukuba, Japan
\item \Idef{org123}University of Zagreb, Zagreb, Croatia
\item \Idef{org124}Universit\'{e} de Lyon, Universit\'{e} Lyon 1, CNRS/IN2P3, IPN-Lyon, Villeurbanne, France
\item \Idef{org125}V.~Fock Institute for Physics, St. Petersburg State University, St. Petersburg, Russia
\item \Idef{org126}Variable Energy Cyclotron Centre, Kolkata, India
\item \Idef{org127}Vestfold University College, Tonsberg, Norway
\item \Idef{org128}Warsaw University of Technology, Warsaw, Poland
\item \Idef{org129}Wayne State University, Detroit, MI, United States
\item \Idef{org130}Wigner Research Centre for Physics, Hungarian Academy of Sciences, Budapest, Hungary
\item \Idef{org131}Yale University, New Haven, CT, United States
\item \Idef{org132}Yonsei University, Seoul, South Korea
\item \Idef{org133}Zentrum f\"{u}r Technologietransfer und Telekommunikation (ZTT), Fachhochschule Worms, Worms, Germany
\end{Authlist}
\endgroup